\documentclass[12pt,pra,aps,amssymb,amsmath,tightenlines,showpacs]{revtex4}
\usepackage{graphicx}

\begin{document}
\title{ Entanglement in the bimodal Jaynes-Cummings model with the
two-mode squeezed vacuum state}
\author{Faisal A. A. El-Orany }
\email{el_orany@yahoo.com}
 \affiliation{ Cyberspace Security
Laboratory, MIMOS Berhad, Technology Park Malaysia, 57000 Kuala
Lumpur, Malaysia;
 Department of Mathematics
and Computer Science, Faculty of Science, Suez Canal University,
Ismailia, Egypt; }
\author{S. Abdel-Khalek }
 \affiliation{ Cyberspace Security
Laboratory, MIMOS Berhad, Technology Park Malaysia, 57000 Kuala
Lumpur, Malaysia}
\author{M. Abdel-Aty }
 \affiliation{ Cyberspace Security
Laboratory, MIMOS Berhad, Technology Park Malaysia, 57000 Kuala
Lumpur, Malaysia}
\author{M. R. B. Wahiddin }
 \affiliation{ Cyberspace Security
Laboratory, MIMOS Berhad, Technology Park Malaysia, 57000 Kuala
Lumpur, Malaysia}

\begin{abstract}
In this paper, we study the interaction between  the two-level
atom and  a bimodal cavity field, namely, two-mode Jaynes-Cummings
model when the atom and the modes are initially in the atomic
superposition state and two-mode squeezed vacuum state,
respectively. For this system we investigate the atomic inversion,
linear entropy and atomic Wehrl entropy. We show that there is a
connection between all these quantities. Also we prove that  the
atomic Wehrl entropy exhibits behaviors similar to those  of the
linear entropy and the von Neumann entropy. Moreover, we show that
the bipartite exhibits periodical disentanglement and derive the
explicit forms of the states of the atom and the modes at these
values of the interaction times.

\end{abstract}
\pacs{42-50.Dv      Jaynes-Cummings model, atomic inversion,
two-mode squeezed vacuum states,  Wehrl entropy, linear entropy.
 } \maketitle

\section{\textbf{Introduction}}
One of the few exactly solvable models in quantum optics, which
describes the interaction between the radiation field and the
matter, is the Jaynes–Cummings model (JCM) \cite{jay1}. The JCM has
become experimentally realizable with the Rydberg atoms in the
high-Q microwave cavities (e.g., see \cite{remp}). Various
extensions for the JCM have been performed and investigated in
greater details such as multiphoton \cite{fa,tw},
intensity-dependent  \cite{buck}, multimode, e.g.,
\cite{fas3,fas2,fas1,chr,two}, multilevel atoms \cite{multil} and
multiatom interactions \cite{multia}. One of these extensions is the
two-mode JCM (TJCM), which has taken a considerable interest in the
literatures, e.g. \cite{fas3,fas2,chr,two,faas,obaad}. The
revival-collapse phenomenon (RCP) occurring in the atomic inversion
of the TJCM is rather complicated compared to that of the standard
JCM (, i.e. a single mode interacting with a single atom) in the
sense that the revival series is compact and each revival is
followed with a secondary revival. Furthermore, the RCP associated
with the TJCM is independent of the initial intensities of the
modes. Such behavior has been partially explained in \cite{chr},
however, an investigation for the occurrence of the secondary
revivals is given in \cite{fas3}. The quantum phase properties for
the TJCM with the initial Schr\"{o}dinger-cat states have been
investigated in \cite{fas2} showing that
 the phase variances of the single-mode
case  can exhibit RCP about the long-time behavior but with the
interaction time several times smaller than that of the standard
JCM. Furthermore,  there is a clear relationship between the RCP
occurring in the atomic inversion  and the behavior of the phase
distribution of both the single-mode and the two-mode cases.
Moreover, for the TJCM it has been shown that there is  a class of
states for which the quadratures squeezing exhibit RCP similar to
that involved in the corresponding atomic inversion \cite{faas}.

Quantum information processing provides different way for
manipulating information than the classical one. This is related
to entanglement, which plays an essential role in the  quantum
information \cite{benn} such as quantum computing \cite{prin},
teleportation \cite{tel1}, cryptographic \cite{cry1}, dense coding
\cite{dens} and entanglement swapping \cite{swap}. Thus
 intensive  efforts have been done to understand theoretically and experimentally
 the entanglement in the quantum systems.
 For instance, the entanglement between two qubits in an
arbitrary pure state has been quantified by  the concurrence
\cite{wott} and Peres-Horodecki measure \cite{peres}, however,
that of the mixed states quantified by the infimum of the average
concurrence over all possible pure state ensemble decomposition.
Additionally, the entropic relations are used in investigating the
entanglement in the quantum system. In this regard von Neuman
entropy (NE) \cite{neum}, linear entropy (LE) and the Shannon
information entropy (SE) \cite{shann} have been frequently used in
treating entanglement in the quantum systems.
 The NE \cite{knig} and the LE \cite{fa} have been
applied to the JCM. It is worth mentioning that  the SE  involves
only the diagonal elements of the density matrix and in some cases
it can give information similar to that obtained from the NE and
LE.
 On the other hand, there is an additional entropic
relation, namely, the Wehrl entropy (WHE) \cite{wehrl}. This
relation has been successfully applied in  description of different
properties of the quantum optical fields such as phase-space
uncertainty \cite{mira1,mira2}, quantum interference \cite{mira2},
decoherence \cite{deco,orl} etc. The WHE is more sensitive in
distinguishing states than the NE since WHE is a state dependent
\cite{mira3}. The concept of the Wehrl phase distribution has been
developed and shown that it serves as a measure of both noise
(phase-space uncertainty) and phase randomization \cite{mira3}.
Furthermore, the WHE has been applied to the dynamical systems, too.
In this respect the time evolution of the WHE for the Kerr-like
medium has been discussed in \cite{jex} showing that the WHE gives a
clear signature for the formation of finite superpositions of
coherent states (cat-like states) as well as
 the number of coherent
components taking part in the superposition. For the JCM the WHE
gives an information on the splitting of the $Q$-function in the
course of the collapse region of the atomic inversion as well as on
the atomic inversion itself \cite{orl,obad}.
 In the present
paper we show that the atomic Wehrl entropy (AHE) \cite{obad} can
be used to quantify entanglement in the TJCM when the modes are
initially prepared in the maximally entangled states such as
 the two-mode
squeezed vacuum state (TMS) \cite{ma}. The TMS can be expressed
as
\begin{equation} |r\rangle =\sum\limits_{n=0}^{\infty}
C_n|n,n\rangle, \label{sv}
\end{equation}
where $C_n=\frac{1}{\cosh r}z^n,\quad z=\tanh r$. We should stress
that the TMSs are important states since they contain quantum
correlations between the different modes that make up the field.
Based on this fact the TMSs have been used in the
continuous-variable teleportation \cite{[7]}, quantum key
distribution \cite{[9]}, verification of EPR correlations
\cite{[12]}, etc. Most importantly the TMSs  can be experimentally
generated via the optical parametric oscillator \cite{[10]}, where
squeezing can be detected with high efficiency homodyne detector
\cite{bich}. Now for the TJCM with the initial TMS  we investigate
the atomic inversion, LE, Wehrl atomic density and AHE. This work
is motivated by the importance of the TJCM  and TMS in quantum
optics as well as in the quantum information, as we mentioned
above. We obtain many of interesting results such as the atomic
Wehrl density (entropy) can give qualitative (quantitative)
information on the entanglement in the TJCM. In other words, the
AWE can be used as a new measure for quantifying entanglement in
the quantum systems. Furthermore, we show that the bipartite (,
i.e. two-mode and atom) can periodically exhibit instantaneous
(long-lived) disentanglement. Also we derive the explicit forms
for the states of the atom and the two-mode  at these values of
the interaction times. The paper is prepared  in the following
order: In section 2 we derive the wavefunction of the system and
evaluate the expectation values of the atomic variables. In
section 3 we investigate the atomic inversion and the LE. In
section 4 we discuss  the atomic Wehrl density and the atomic
Wehrl entropy. In section 5 we summarize the main results.

\section{\textbf{Wavefunction of the system}}
In this section we derive the dynamical wavefunction for the
system, which consists of two-mode field interacting with the
two-level atom. Also, we evaluate the expectation values for the
atomic operators which will be frequently used in the paper.

The Hamiltonian describing the interaction between two-mode field
and the two-level atom in the rotating wave approximation takes
the form \cite{sh}:

\begin{equation}
\hat{H}=\hbar (\hat{H}_0+\hat{H}_{i}),\label{f1}
\end{equation}
where
\begin{eqnarray}
\begin{array}{lr}
\hat{H}_0=\sum\limits_{j=1}^{2} \omega _{j}\hat{a}_{j}^{\dagger
}\hat{a}
_{j} +\omega _{a}S_{z},\\
\\
\hat{H}_i=\chi (\hat{a}_{1}^{\dagger 2}\hat{a}_{1}^{2}+
\hat{a}_{2}^{\dagger 2}\hat{a}_{2}^{2})+
 \lambda \left[ \hat{a}_{1}\hat{a}_{2}\hat{S}_{+}+
 \hat{a}^{\dagger}_{1}\hat{a}^{\dagger}_{2}
\hat{S}_{-} \right] ,  \label{5-1}
\end{array}
\end{eqnarray}
where  $\hat{a}_{j}\quad (\hat{a}_{j}^{\dagger })$ and $\omega_j$
are the annihilation (creation) operator and frequency designated
 the $j$th mode, the set $\{
\hat{S}_{z},\hat{S}_{+},\hat{S}_{-}\}$ is the usual Pauli spin
operators, $\lambda $ is the coupling constant between the
 atom and the modes, $\chi$
is the dispersive part of the third-order nonlinearity of the
Kerr-like medium, which is  assumed to be  the same for the two
modes. Also  $\omega _{a}$ denotes the atomic transition
frequency.

Throughout the investigation we consider that  $\omega _{a}=\omega
_{1}+\omega _{2}$ (i.e. the resonance case), the modes are
initially prepared in the TMS (\ref{sv}) and the atom is in the
atomic superposition state having the form:
\begin{equation}
\mid \Psi _{a}(0)\rangle =\cos (\frac{\theta}{2})\mid \uparrow
\rangle +\sin (\frac{\theta}{2})\exp(i\phi) \mid \downarrow
\rangle , \label{5-super}
\end{equation}%
where $\mid \uparrow \rangle$ and $\mid \downarrow \rangle$ deonte
excited and ground atomic states, respectively; $\theta$ and
$\phi$ are phases. It is worth reminding that
 preparing the
atom in the coherent superposition states is important as a result
of its applications in noise quenching by correlated spontaneous
emission \cite{spa1}, quantum beats \cite{spa2} and noise-free
amplification \cite{spa3}. Now under these conditions
 the dynamical wavefunction for the Hamiltonian
system (\ref{5-1}) can be evaluated in the standard way as
\begin{equation}
\mid \Psi (T)\rangle =\sum\limits_{n=0}^{\infty}\exp(-2i\eta Tn^2)
\left[ F_1(n,T)\mid \uparrow,n,n \rangle +F_2(n,T) \mid \downarrow,
n+1,n+1 \rangle\right] , \label{5f}
\end{equation}
where
\begin{eqnarray}
\begin{array}{lr}
F_1(n,T)=C_n \cos\frac{\theta}{2}\cos(T\Omega_n)+\frac{i
C_n}{\Omega_n}\left[ 2\eta n \cos\frac{\theta}{2}
-(n+1)\exp(i\phi)\sin\frac{\theta}{2}\right]\sin(T\Omega_n),\\
\\
F_2(n,T)=C_n
\exp(i\phi)\sin\frac{\theta}{2}\cos(T\Omega_n)-\frac{i
C_n}{\Omega_n}\left[ 2\eta n \exp(i\phi)\sin\frac{\theta}{2}
+(n+1)\cos\frac{\theta}{2}\right]\sin(T\Omega_n) \label{5f1}
\end{array}
\end{eqnarray}
and
\begin{equation}
\Omega_n=\sqrt{4\eta^2 n^2+(n+1)^2},\quad
\eta=\frac{\chi}{\lambda},\quad T=t\lambda.\label{rr}
\end{equation}
 All information about the system is
involved in  the wavefunction (\ref{5f}) or in the total density
matrix $\hat{\rho}(T)=\mid \Psi (T)\rangle \langle \Psi (T)|$. As
we mainly deal with the atomic subsystem  we evaluate
 the atomic reduced
 density matrix $\hat{\rho}_a(T)$ via the relation:
\begin{equation}
\hat{\rho}_a(T)={\rm Tr}_f\hat{\rho}(T), \label{ff1}
\end{equation}
where the subscript $f$ means that the trace is taken over the
field. We close this section by evaluating the expectation values
for  the atomic operators $\{
\hat{S}_{z},\hat{S}_{x},\hat{S}_{y}\}$ for (\ref{5f}) as
\begin{eqnarray}
\begin{array}{lr}
\langle \hat{S}_{x}(T)\rangle=\frac{1}{2}\rho _{x}(T) ={\rm Re}
\sum\limits_{n=0}^{\infty}
\exp[2i\eta T(2n+1)] F_1^{*}(n+1,T)F_2(n,T),\\
\\
\langle \hat{S}_{y}(T)\rangle=\frac{1}{2}\rho _{y}(T) ={\rm Im}
\sum\limits_{n=0}^{\infty}
\exp[2i\eta T(2n+1)] F_1^{*}(n+1,T)F_2(n,T),\\
\\
\langle \hat{S}_{z}(T)\rangle=\frac{1}{2}\rho _{z}(T) =
\frac{1}{2}\sum\limits_{n=0}^{\infty} [
|F_1(n,T)|^2-|F_2(n,T)|^2],\label{fba}
\end{array}
\end{eqnarray}%
where $\hat{S}_{x}=\frac{1}{2}(\hat{S}_{-}+\hat{S}_+),
\hat{S}_{y}=\frac{1}{2i}(\hat{S}_{+}-\hat{S}_{-})$.
 We use the results obtained in this section to make a
comparative study between the atomic inversion, LE and AWE in the
following sections.

\section{\textbf{Atomic inversion and linear entropy }}
In this section we investigate the atomic inversion and the LE for
the system under consideration. We start the investigation with
the atomic inversion, which is plotted in Fig. 1  versus the
scaled time $T$ for the given values of the interaction
parameters. From this figure the obvious remark is that the
$\langle \hat{S}_{z}(T)\rangle$ is periodic with period $\pi$,
regardless of the type of the initial atomic state.
 The origin in this
 is related to the nature of the TMS
in which the information is equally distributed among the two
modes \cite{mil}. Furthermore, such periodicity is quite similar
to that of the two-photon JCM \cite{tw} and intensity-dependent
JCM \cite{buck}. For the superposition atomic state
$(\theta,\phi)=(\pi/2,\pi/2)$, i.e. the dashed curve, one can
observe that $\langle \hat{S}_{z}(T)\rangle$ oscillates around
zero with extreme values $\pm 0.15$.
 Nevertheless, we have
found that the system  can exhibit atomic trapping, i.e. $\langle
\hat{S}_{z}(T)\rangle=0$, for $(\theta,\phi)=(\pi/2,0)$, as we
show shortly.
 All these facts can be analytically realized for
$\eta=0$. In this case $\langle \hat{S}_{z}(T)\rangle$ in
(\ref{fba}) can be easily modified as:

%%%%%%%%%%%%%%%%%%%%%%%%%%%%%%%%%%%%%%%%%%%%%%%%%%%%%%%%%%%%%%%
\begin{figure}
  \vspace{0cm}
    \includegraphics[width=0.86\linewidth]{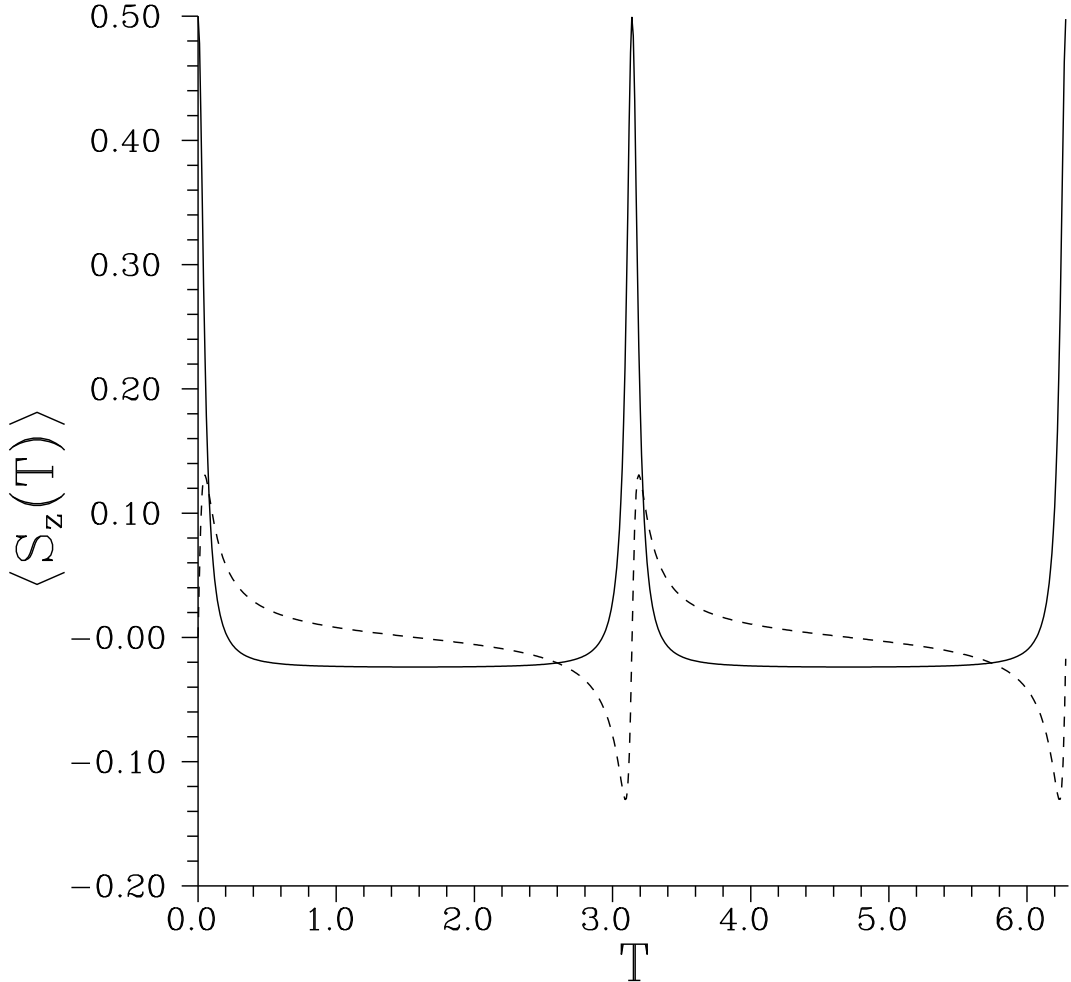}
\caption{ The $\langle \hat{S}_{z}(T)\rangle$  versus scaled time
$T$
 for
$\sinh^2 r=10$ and $(\theta,\phi,\eta)=(0,0,0)$ (solid curve) and
$(\pi/2,\pi/6,0)$ (dashed curve).}
\end{figure}
%%%%%%%%%%%%%%%%%%%%%%%%%%%%%%%%%%%%%%%%%%%%%%%%%%%%%%%%%%%%

\begin{equation}
\langle \hat{S}_{z}(T)\rangle=\frac{[\cos(2T)-z^2]\cos\theta +
z\sin\theta\sin(\phi)\sin(2T)}
{2[(1-z^2\cos(2T))^2+z^4\sin^2(2T)]\cosh^2 r}. \label{fIaI}
\end{equation}
 From (\ref{fIaI})
the smallest value of the atomic inversion for $\theta=0$ is
$\langle \hat{S}_{z}(T)\rangle=-1/(2+2\bar{n})$, where
$\bar{n}=2\sinh^2 r$, which occurs at
 $T=\pi/2$ (, i.e. collapse time). This means  that the greater the initial
mean-photon number of the field the closer the value of the
$\langle \hat{S}_{z}(T)\rangle$ to zero at the collapse time. Also
it is obvious that
 for $(\theta,\phi)=(\pi/2,0)$ the expression (\ref{fIaI}) tends to zero.
We have found that the behavior of the $\langle
\hat{S}_{z}(T)\rangle$ is
 insensitive of the value of $\eta$. Finally
the information obtained here is important in investigating the
entanglement in the bipartite, as we show below.

Now we quantify the entanglement between the two-mode system and
the atomic system  via the linear entropy, which is defined by
\begin{equation}
\xi (T)=1-{\rm Tr}\hat{\rho}^2_a(T). \label{sIaI}
\end{equation}
As a result of the symmetry in the system the linear entropies for
the atomic and field sub-systems are equivalent.  Furthermore, as
the system is in an over all pure state, the LE is a relevant
measure for quantifying entanglement in the bipartite \cite{rung}.
 The $\xi (T)$ ranges from $0$ for pure states and/or
disentangled states to $0.5$ for a maximally entangled states
\cite{julio}. For the future purpose, the relation (\ref{sIaI})
can be expressed in terms of the expectation values of the atomic
variables (\ref{fba}) as
\begin{equation}
\xi(T)=\frac{1}{2}\left[1-\rho^2 _{x}(T)-\rho^2 _{y}(T)-\rho^2
_{z}(T)\right]. \label{avaI}
\end{equation}
 In Figs. 2 we plot $\xi(T)$
versus the scaled time $T$ for given values of the interaction
parameters. From Fig. 2(a) and Fig. 2(b), where $\eta=0$, the
bipartite is periodically disentangled with period $\pi$. As we
mentioned above this is related to the nature of the TMS. Also
this is manifested in the wavefunction (\ref{5f}) as linear Rabi
frequency $\Omega_n$ in the parameter $n$.  Moreover, from Figs.
2(a)-(b) the
 $\xi(T)= 0$ at $T=0$, and when the
interaction is switched on, i.e. the interchange of energy between
the two modes and atom occurs, the  bipartite exhibits
  immediately maximum  entanglement (i.e. $\xi(T)$ provides maximum value),
stays for a while then exhibiting long-lived disentanglement at
$T=\pi/2$. As the interaction proceeds, the $\xi(T)$ goes
 gradually to the maximum value (i.e. entanglement) and
eventually tends to zero  at $T=\pi$. Furthermore, the comparison
between Fig. 2(a) and Fig. 2(b) shows that the atomic coherence
decreases the amount of entanglement in the bipartite. This agrees
with the fact that the interference in phase space decreases the
degree of entanglement in the bipartite \cite{fais1}. To obtain
more accurate information on the disentanglement in the bipartite
for $\eta=0$ we give a closer look at the analytical form of the
wavefunction (\ref{5f}) at $T=m\pi$ and $T=m'\pi/2$, where $m$ and
$m'$ are integer and odd integer numbers, respectively. At these
values of the interaction times  the wavefunction (\ref{5f}) can
be easily expressed as:

%%%%%%%%%%%%%%%%%%%%%%%%%%%%%%%%%%%%%%%%%%%%%%%%%%%%%%%%%%%%%%%
\begin{figure}
  \vspace{0cm}
    \includegraphics[width=0.86\linewidth]{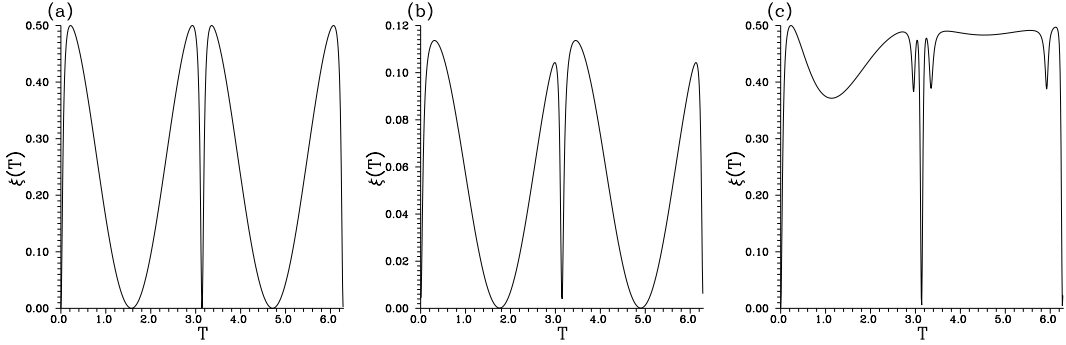}
\caption{ The $\xi(T)$  versus scaled time $T$
 for
$\sinh^2 r=10$ and $(\theta,\phi,\eta)=(0,0,0)$ (a),
$(\pi/2,\pi/6,0)$ (b) and $(0,0,0.03)$ (c).}
\end{figure}
%%%%%%%%%%%%%%%%%%%%%%%%%%%%%%%%%%%%%%%%%%%%%%%%%%%%%%%%%%%%

\begin{equation}
 |\psi (T=m\pi)\rangle
=\sum\limits_{n=0}^{\infty}C_n\exp[im(n+1)\pi]|n,n\rangle \otimes
\Bigl[\cos(\frac{\theta}{2}) |\uparrow \rangle
+\sin(\frac{\theta}{2}) \exp[i(\phi-m\pi)] |\downarrow\rangle
\Bigr], \label{sIaI1}
\end{equation}

\begin{eqnarray}
\begin{array}{lr}
|\psi (T=\frac{m'}{2}\pi)\rangle
=\sum\limits_{n=0}^{\infty}\exp(in\pi)\Bigl[
C_{2n}\exp(i\phi)\sin(\frac{\theta}{2})|2n,2n\rangle\\
\\
-
C_{2n+1}\exp(i\frac{m'\pi}{2})\cos(\frac{\theta}{2})|2n+1,2n+1\rangle\Bigr]
\otimes [\exp(-i\frac{m'\pi}{2}) |\uparrow\rangle +z^{-1}
|\downarrow\rangle], \label{sIaI2}
\end{array}
\end{eqnarray}
From (\ref{sIaI1}) it is evident that when $T=m\pi$ the bipartite
tends to its initial form, but the states of the field and atom
include additional phase factors.   This behavior is quite similar
to that of the standard JCM \cite{fa} and two-atom JCM
\cite{fais1}. Nevertheless, from (\ref{sIaI2}) the bipartite is
disentangled, however, some information from the field is involved
in the  atomic system and vice versa. It seems that this is the
origin in the long-lived disentanglement at these values of the
interaction times. Moreover, expression (\ref{sIaI2}) shows that
when the atom is initially in the excited (ground) state the
system can generate odd (even) two-mode squeezed vacuum state.
Thus at one-half revival time the system can generate two-mode cat
states, which is similar to the standard JCM \cite{fa}. We
conclude this part by writing down the explicit form of the
wavefunction for $(\theta,\phi,\eta)=(\pi/2,0,0)$, which can be
evaluated  as

\begin{equation}
 |\psi (T)\rangle
=\frac{1}{\sqrt{2}}\sum\limits_{n=0}^{\infty}C_n\exp[i(n+1)T]|n,n\rangle
\otimes \Bigl[ |\uparrow \rangle +
 |\downarrow\rangle \Bigr]. \label{csIaI1}
\end{equation}
From (\ref{csIaI1}) it is evident that the bipartite exhibits
always disentanglement, i.e. $\xi(T)=0$. From the above
investigation one can notice that for $\eta=0$  the behavior of
the $\xi (T)$ is consistent with that of the atomic inversion.

Now we draw the attention to the influence of the Kerr-like medium
on the behavior of the $\xi(T)$ for the system under consideration
(see Fig. 2(c)). The comparison between Fig. 2(a) and Fig. 2(c)
shows that involving Kerr-like medium in the system drastically
changes the behavior of the $\xi(T)$, e.g. it increases the amount
of entanglement in the bipartite. This is related to several
facts:
 (i) Kerr-like medium provides  non-trivial photon-dependent phase in
  the wavefunction (cf. (\ref{5f})) as well as it  makes
the Rabi oscillation as a nonlinear function in terms of  $n$.
 (ii) The Kerr-like medium alone  has a nonclassical nature where it
 can generate superposition states such as Yurke-Stoler states
\cite{[45]}.  It is obvious that for $\eta\neq 0$ the evolutions
of the $\langle \hat{S}_{z}(T)\rangle$ and $\xi (T)$ generally are
 not consistent.

The final remark, for the system under consideration  we have
found that the  von Neumann entropy gives typical behavior as that
of the $\xi(T)$. In the following section we study the evolution
of the atomic Wehrl entropy and compare it with that of the
$\xi(T)$.

\section{\textbf{Atomic Wehrl density and entropy}\textbf{\ }}
In this section we investigate the atomic Wehrl density and atomic
Wehrl entropy AWE for obtaining  more information on the  system.
Also we compare the behavior of these quantities with those given
in section 3. We start the investigation by defining the atomic
$Q$-function as \cite{karol}:
\begin{equation}
Q_{a}(\Theta ,\Phi ,T)=\frac{1}{2\pi }\left\langle \Theta ,\Phi
\left\vert \hat{\rho}_{a}(T)\right\vert \Theta ,\Phi \right\rangle
,  \label{aw1}
\end{equation}%
where $\hat{\rho}_{_a}(T)$ is the atomic reduced density matrix
(\ref{ff1}) and $\left\vert \Theta ,\Phi \right\rangle $ is the
atomic coherent state expressed as
\begin{equation}
\left\vert \Theta ,\Phi \right\rangle =\cos \left( \Theta
/2\right) \left\vert \uparrow \right\rangle +\sin \left( \Theta
/2\right) e^{i\Phi }\left\vert \downarrow \right\rangle,
\label{aw3}
\end{equation}
where $0\leq \Theta\leq \pi, 0\leq \Phi \leq 2\pi$.  The
definition (\ref{aw1}) means that two different spin coherent
states overlap unless they directed into two antipodal points on
the sphere \cite{karol}. Here we shows that $Q_{a}$ can give
qualitative information on the entanglement in the bipartite.
 From (\ref{ff1}),
(\ref{aw1}) and (\ref{aw3}) one can evaluate $Q_{a}(\Theta ,\Phi
,T)$ as

\begin{equation}
Q_{a}(\Theta ,\Phi ,T)=\frac{1}{4\pi }\left\{ 1+\rho _{z}(T)\cos
\Theta +\left[ \rho _{x}(T)\cos \Phi +\rho _{y}(T)\sin \Phi
\right] \sin \Theta \right\}.  \label{aw4}
\end{equation}%
One can easily check that the $Q_{a}$ given by (\ref{aw4}) is
normalized. From (\ref{aw4}) it is obvious that  at particular
values of the $\Theta$ and $\Phi$ the $Q_{a}$ can give information
on the dipole moment and the atomic energy. Moreover, we have
found that the behavior of the $Q_{a}(\Theta ,\Phi ,T)$ confirms
the information obtained in the previous section. For instance,
when the atom is initially in the excited or ground state and
$T=m\pi$, where $m$ is integer, the $Q_{a}(\Theta ,\Phi ,T)$
exhibits behavior as that of the initial case. More
illustratively, at these values of the interaction times we have
$\rho _{z}(T)=\rho _{z}(0), \quad \rho _{x}(T)=\rho _{y}(T)=0$ and
hence (\ref{aw4}) reduces to
\begin{equation}
Q_{a}(\Theta ,\Phi ,T=m\pi)=\frac{1}{4\pi }\left[ 1+\rho
_{z}(0)\cos \Theta
 \right] .  \label{aw4aa}
\end{equation}%
From (\ref{aw4aa}),  $Q_{a}$ is independent of $\Phi$ and hence it
exhibits wave shape in the $\Theta-\Phi$ plane with amplitude
$\rho _{z}(0)$. On the other hand, when the atom is in the excited
or ground or superposition states with
$(\theta,\phi)=(\pi/2,\pi/2)$ and $T=m\pi/2$, where $m$ is odd
integer, the $Q_{a}(\Theta ,\Phi ,T)$ exhibits two-peak structure
 (, i.e. one peak is up and the other is down) as shown in
  Fig. 3a (for given values of the interaction
parameters). Additionally, we plot $Q_{a}(\Theta ,\Phi ,T)$ for
the atomic superposition state at $T=\pi$ in Fig. 3(b). The
comparison between Fig. 3(a) and Fig. 3(b) shows that the
locations of the maximum and minimum values in the $Q_{a}(\Theta
,\Phi ,T)$ are interchanged even though in these cases the
bipartite is disentangled (cf. (\ref{sIaI2})). For atomic trapped
case (cf. (\ref{csIaI1})) we have $ \rho _{x}(T)=\rho _{x}(0),
\quad \rho _{y}(T)=\rho _{z}(T)=0$ and hence (\ref{aw4}) becomes
time independent.
 Now we draw the attention to the influence of the
 $\eta$ on the behavior of the $Q_{a}(\Theta ,\Phi ,T)$, which is
 plotted  in Fig. 3(c).
 We have generally found when the atom is in the excited or ground state
 and $T=m\pi$  the $Q_{a}(\Theta ,\Phi ,T)$ is
 insensitive of $\eta$, however, at $T=m\pi/2$,
 e.g. for the excited atomic state,
the two-peak structure involved in the $Q_{a}$ for $\eta=0$  is
destroyed (compare Fig. 3(b) and Fig. 3(c)). This confirms that
the bipartite is entangled (disentangled) at $T=\pi/2$ ($T=\pi$).
 From the facts presented above and previous section
we can conclude that $Q_a$ can give qualitative information on the
entanglement in the bipartite.  It is worth mentioning that the
$Q$ function of the field can manifest squeezing involved in the
field components, however, this is not the case for $Q_a$.
 This is related
to the difference between the bosonic and fermonic algebras.

%%%%%%%%%%%%%%%%%%%%%%%%%%%%%%%%%%%%%%%%%%%%%%%%%%%%%%%%%%%%%%%
\begin{figure}
  \vspace{0cm}
    \includegraphics[width=0.86\linewidth]{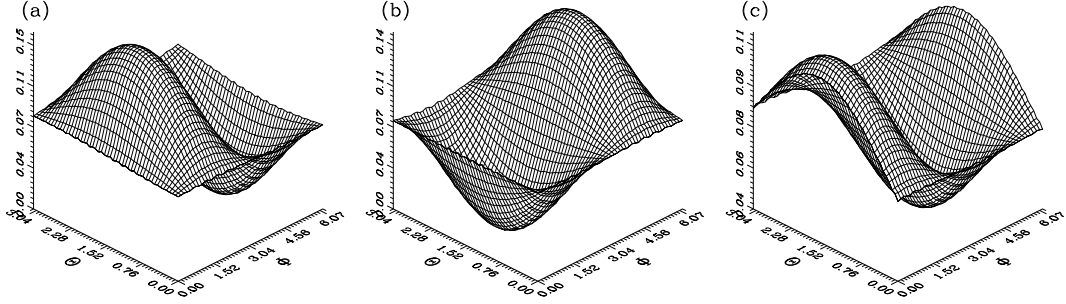}
\caption{ The $Q_{a}(\Theta ,\Phi ,T)$ for $\sinh^2 r=10$ and
$(\theta,\phi,T,\eta)=(0,0,\pi/2,0)$ (a), $(\pi/2,\pi/2,\pi,0)$ (b)
and $(0,0,\pi/2,0.03)$ (c).}
\end{figure}
%%%%%%%%%%%%%%%%%%%%%%%%%%%%%%%%%%%%%%%%%%%%%%%%%%%%%%%%%%%%

Now we draw the attention to the AWE, which is defined as
\cite{obad}:
\begin{equation}
W_{a}(T)=-\int_{0}^{2\pi }\int_{0}^{\pi } Q_{a}(\Theta ,\Phi
,T)\ln Q_{a}(\Theta ,\Phi ,T)\sin \Theta d\Theta d\Phi .
\label{aw8}
\end{equation}%
It is worth reminding that the definition (\ref{aw8}) is given in
analogy with that of the field Wehrl entropy \cite{wehrl}.
Furthermore, the $W_{a}(T)$ cannot be negative as a result of the
$Q_a$ is a non-negative function.  As it is generally difficult to
find a closed form  for the $W_{a}(T)$ numerical techniques have
to be used. Nevertheless, at particular values of the interaction
parameters  the exact form can be obtained. For instance, for
(\ref{aw4aa}) the relation (\ref{aw8})
 can be evaluated as
\begin{equation}
W_{a}(T=m\pi) = \frac{1}{2}+{\rm ln}(4\pi)+\frac{1+\rho
_{z}^2(0)}{4\rho _{z}(0)}{\rm \ln} \left[\frac{1-\rho _{z}(0)}
{1+\rho _{z}(0)}\right] -\frac{1}{2}{\rm \ln} \left(1-\rho
_{z}^2(0)\right).
  \label{aw10}
 \end{equation}
From expression (\ref{aw10}) for atom initially in the excited or
ground state we obtain  $W_a(T=m\pi)\simeq 2.488$.
 In the language of entanglement the bipartite is disentangled when
$W_a\simeq 2.488$.
 This value represents  the lower
bound of the $W_{a}(T)$.  In Figs. 4(a) and (b) we plot $W_{a}(T)$
for the same values of the interaction parameters as in Figs. 2(a)
and (c), respectively. Surprisingly one can observe that the
behaviors in the Figs. 4(a) and (b) are completely similar to
those in the Figs. 2(a) and (c), respectively. We have checked
this fact for different values of the interaction parameters and
obtained the same conclusion. The similarity between the behaviors
of the $\xi (T)$ and $W_a(T)$  can be  realized from the relations
(\ref{avaI}) and (\ref{aw4}), i.e. the LE and the atomic Wehrl
density and/or AWE are function in $ \rho _{x}(T), \rho
_{y}(T),\rho _{z}(T)$.
 Moreover, the upper bound of the AWE can be
obtained by comparing the maximum value of the LE $(=0.5)$  with
the maximum value of the AWE $(\simeq 2.53102)$. Thus the upper
bound of the AWE is $2.53102$. It is obvious that the AWE is
compact informative one-parameter measure describing the
entanglement dynamics of the system at all values of the
interaction time.
 From this information we can conclude that the $W_a(T)$ is
  a new measure for quantifying the
entanglement in the bipartite.
%%%%%%%%%%%%%%%%%%%%%%%%%%%%%%%%%%%%%%%%%%%%%%%%%%%%%%%%%%%%%%%
\begin{figure}
  \vspace{0cm}
    \includegraphics[width=0.86\linewidth]{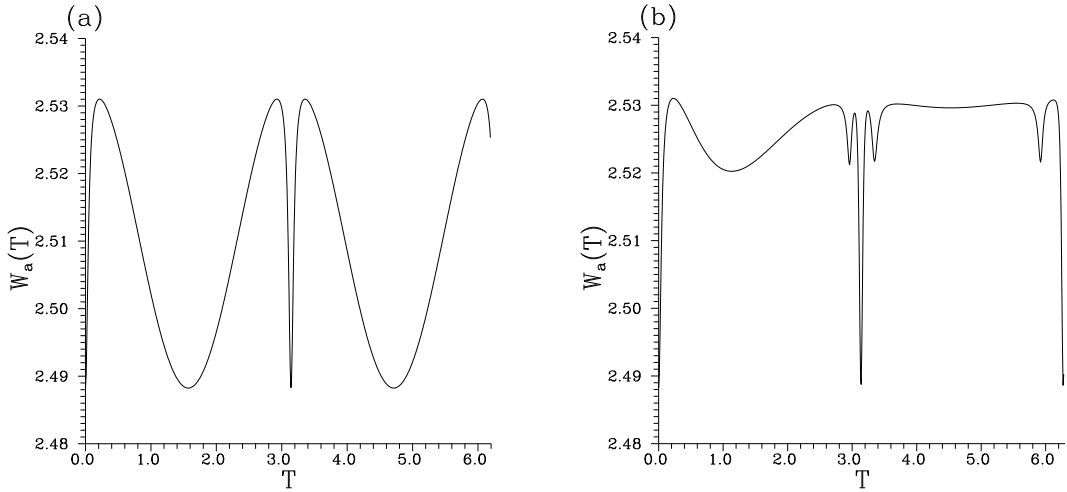}
\caption{ The $W_a(T)$  versus scaled time $T$
 for
$\sinh^2 r=10$ and $(\theta,\phi,\eta)=(0,0,0)$ (a) and $(0,0,0.03)$
(b).}
\end{figure}
%%%%%%%%%%%%%%%%%%%%%%%%%%%%%%%%%%%%%%%%%%%%%%%%%%%%%%%%%%%%

The final remark is: the LE associated with the reduced atomic and
reduced field density matrices  can give typical information on
the entanglement in the bipartite. This is not the case for the
field Wehrl entropy $W_f$ and the atomic Wehrl entropy $W_a$. For
instance, as we mentioned in the Introduction the $W_f$ gives
information on the RCP occurred in the atomic inversion as well as
on the splitting of the Husimi $Q$ function in the course of the
collapse region \cite{orl}. The difference between the behaviors
of the $W_f$ and $W_a$ is related to that the $Q$ functions of the
field and atom provide
 different information on the system, where the former and the
 latter give information on the atomic inversion \cite{ris1} and on the
entanglement in the bipartite, respectively.

\section{Conclusion}
In this paper we have studied the properties of the TJCM
(involving a Kerr-like medium) when the modes and atom are
initially in the TMS and atomic superposition state, respectively.
 Particularly, we have investigated the
atomic inversion, LE, atomic Wehrl density and AWE. We have shown
that all these quantities can give information on the entanglement
in the bipartite. The bipartite exhibits instantaneous and
long-lived disentanglement at particular values of the interaction
parameters. We have derived the explicit forms for the atomic and
field states at the disentanglement times. We have shown that the
atomic Wehrl entropy (density) gives quantitative (qualitative)
information on the entanglement in the bipartite. Most
importantly, the LE and AWE can give typical information on the
entanglement in the bipartite. This is related to that these
quantities can be expressed in terms of the expectation values of
the atomic operators.
 In this regard the
$W_a$ can be interpreted as an information measure for such
system.
 Finally, we have shown that the Kerr-like medium
increases the degree of entanglement in the bipartite.

%%%%%%%%%%%%%%%%%%%%%%%%%%%%%%%%%%%%%%%%%%%%%%%%%%%%%%%%%%%%%
\section*{ Acknowledgement}
%%%%%%%%%%%%%%%%%%%%%%%%%%%%%%%%%%%%%%%%%%%%%%%%%%%%%%%%%%%%%%

The authors (F.A.A.E., S. A-K,  M. A-A) are grateful to the
 Cyberspace Security Laboratory, MIMOS Berhad,
Technology Park Malaysia, 57000 Kuala Lumpur, Malaysia for
hospitality and financial support.

\end{document}